\documentclass[11pt]{article}
\usepackage[a4paper,bottom=4.2cm,top=1cm,head=3cm,width=18cm,dvipdfm]{geometry}
\evensidemargin=\oddsidemargin

\usepackage[dotinlabels]{titletoc}
\usepackage{titlesec}
\usepackage[numbers,sort&compress]{natbib} 

\usepackage{xcolor}
\usepackage{mathrsfs}
\usepackage{amsmath}
\usepackage{amssymb}
\usepackage{bm}
\usepackage{amsfonts}
\usepackage{subfigure}
\usepackage{feynmp}
\usepackage{extarrows}
\usepackage{slashed}
\usepackage{graphicx}
\usepackage{dcolumn}
\usepackage{verbatim}
\usepackage{here}
\usepackage{multirow} 
\allowdisplaybreaks[4]
\usepackage{indentfirst}
\usepackage{isodateo}
\usepackage{enumerate}
\usepackage[low-sup]{subdepth}

\numberwithin{equation}{section}
\renewcommand{\thefootnote}{\arabic{footnote}}
\usepackage{isodateo}
\usepackage[CJKbookmarks=true, bookmarksnumbered=true,bookmarksopen=true,]{hyperref}
\hypersetup{colorlinks,%
	linkcolor=blue,
	citecolor=blue,
urlcolor=blue}

\graphicspath{{figs/}}

\newcommand{\be}{\begin{equation}}
\newcommand{\ee}{\end{equation}}
\newcommand{\bea}{\begin{eqnarray}}
\newcommand{\eea}{\end{eqnarray}}

\def\ede{\end{equation}}
\def\bga{\begin{aligned}}
\def\eda{\end{aligned}}
\newcommand{\beq}{\begin{equation}}
\newcommand{\eeq}{\end{equation}}
\newcommand{\bq}{\begin{equation}}
\newcommand{\eq}{\end{equation}}
\newcommand{\ba}{\begin{array}}
\newcommand{\ea}{\end{array}}
\newcommand{\beqa}{\begin{eqnarray}}
\newcommand{\eeqa}{\end{eqnarray}}
\newcommand{\beqs}{\begin{subequations}}
\newcommand{\eeqs}{\end{subequations}}

\def\({\left(}
\def\){\right)}

\def\End{\end{document}}

\def\be{\beta}

\def\End{\end{document}}

\begin{document}

 \thispagestyle{empty}
 \renewcommand{\thefootnote}{\fnsymbol{footnote}}
 \setcounter{footnote}{0}
 \titlelabel{\thetitle.\quad \hspace{-0.8em}}
\titlecontents{section}
              [1.5em]
              {\vspace{4mm} \large \bf}
              {\contentslabel{1em}}
              {\hspace*{-1em}}
              {\titlerule*[.5pc]{.}\contentspage}
\titlecontents{subsection}
              [3.5em]
              {\vspace{2mm}}
              {\contentslabel{1.8em}}
              {\hspace*{.3em}}
              {\titlerule*[.5pc]{.}\contentspage}
\titlecontents{subsubsection}
              [5.5em]
              {\vspace{2mm}}
              {\contentslabel{2.5em}}
              {\hspace*{.3em}}
              {\titlerule*[.5pc]{.}\contentspage}
\titlecontents{appendix}
              [1.5em]
              {\vspace{4mm} \large \bf}
              {\contentslabel{1em}}
              {\hspace*{-1em}}
              {\titlerule*[.5pc]{.}\contentspage}


\vspace*{8mm}

\begin{center}
	{\Large\bf Inelastic dark matter mediated by natural dark photon}
	
\vspace*{8mm}
	
{\large\sc Jie Tang},\footnote{Email: tangj@seu.edu.cn}~
{\large\sc Pei-Hong Gu},\footnote{Email: phgu@seu.edu.cn}~

\vspace*{4mm}
	
School of Physics, Jiulonghu Campus, Southeast University, Nanjing 211189, China

\vspace*{20mm}
\end{center}

\vspace*{3mm}

\begin{abstract}
\baselineskip 17pt   
\noindent

At loop level, a dark $U(1)_X^{}$ gauge field for dark photon can naturally acquire a tiny kinetic mixing with the standard model $U(1)_Y^{}$ gauge field for hypercharge after a $Z_2^{}$ discrete symmetry is spontaneously broken for generating a small mass difference between two vector-like fermions carrying the same $U(1)_Y^{}$ charge and the opposite $U(1)_X^{}$ charges of equal magnitude. To avoid an unacceptable relic density with electric charge, these vector-like fermions can decay into the standard model charged leptons with a TeV-scale complex dark scalar. The real and imaginary parts of this dark scalar can elegantly obtain a mass split below the MeV scale after a dark Higgs scalar develops its vacuum expectation value for generating a dark photon mass of the GeV order. Therefore, the dark photon can mediate an exothermic inelastic scattering of dark matter off nucleus to naturally explain the recently reported event in the LUX-ZEPLIN experiment and substantially relax the stringent constraints from the IceCube neutrino searches. 

\end{abstract}

  \newpage
\renewcommand{\thefootnote}{\arabic{footnote}}
\setcounter{footnote}{0}
\setcounter{page}{2}

\tableofcontents

\setcounter{footnote}{0}
\renewcommand{\thefootnote}{\arabic{footnote}}

\baselineskip 18pt

\vspace*{10mm}
\section{Introduction}
\vspace*{1.5mm}
\label{sec:intro}
\label{sec:1}

Recently the LUX-ZEPLIN (LZ) collaboration has reported a single event at nuclear recoil energy $248 \pm 23 (\textrm{stat}) \pm 23 (\textrm{sys}) \,\textrm{keV}$ in a region with a low expected background \cite{lz2026}. The analysis finds a local significance of $3.4\,\sigma$ and a global significance of $2.6\,\sigma$. The high recoil energy and the absence of a corresponding low-energy excess should imply a dark matter interaction beyond the usual elastic spin-independent limit, if this event is really from the dark matter scattering \cite{mauro2026,visinelli2026,jssgjfb2026,jbdjln2026,chdl2026,myqfwystyzf2026,noos2026,wagkl2026,xydwhkx2026,dbsksnssh2026,gwybzwyc2026,pzgdvgxgw2026,vshllr2026,sjlty2026,langho2026,acddsapapaprp2026,fhwz2026,xqhs2026,rkhkp2026,hmlee2026,xdfw2026,nlrbrstrswlx2026,cmcabe2026,junwin2026,ggllsstyx2026,hbvb2026,lwyx2026,wkkkems2026,jhlzlvqtyx2026,mdmhs2026,pdbksmpkp2026,hadkkkjcpss2026,sbjcfl2026,paabpjfsdhgdk2026,ikscgmaacy2026,yhe2026,dfsvk2026,dbdbpb2026,sjam2026,kkmras2026,mhnz2026,acamdkprs2026,smpkp2026-2,pbsmnnpkp2026,bbarman2026,agicck2026,alyfr2026,adtn2026-2,waaamur2026,hghxhsj2026,ljjmy2026,hbvb2026-2,pujjrp2026,spmtat2026,sfgotyw2026,gammadfq2026,hafgjlmlcx2026,nntty2026,mdm2026-2,nodr2026-3,cajoh2026,bsrml2026,fsjt2026,bcss2026-2,jskz2026-2,bde2026,wagkl2026-2,kcmbaac2026,hjscp2026,cgdhgk2026,ddsk2026-2}. Actually, the so-called inelastic dark matter scattering has motivated several interpretations of the LZ event. However, many well-motivated models for inelastic dark matter scatterings, such as the Higgsino in supersymmetry, may be in tension with other observational constraints \cite{mphr2026}. Specifically, the dark matter particle accelerates in the solar potential so that it can scatter inelastically on heavy nuclei. This solar capture followed by annihilations into electroweak states may produce a detectable neutrino flux. Therefore, the absence of the corresponding IceCube signal excludes the thermal Higgsino-like interpretation of the LZ event \cite{mphr2026,dbose2026,ttqntldh2026}.

On the other hand, the dark photon \cite{okun1982,holdom1986,fh1991} from an artificially introduced $U(1)_X^{}$ gauge group is a hypothetical particle with very attractive phenomena and has been extensively studied by many theorists and experimentalists \cite{fgl2005,cmohv2021,acre2026}. The interactions of the dark photon to the standard model (SM) are assumed from an Abelian kinetic mixing between the SM $U(1)_Y^{}$ gauge field and the new $U(1)_X^{}$ gauge field. If the $U(1)_X^{}$ dark photon is not heavy enough, the $U(1)_X^{}\times U(1)_Y^{}$ kinetic mixing should be small enough to satisfy the stringent constraints from different experiments including low-energy colliders, meson decays, beam dump experiments, high-energy colliders and so on \cite{takahashi2026}. However, the smallness of a renormalizable $U(1)$ kinetic mixing can not be taken for granted.

In this paper we shall construct a $Z_2^{}$ discrete symmetry to naturally generate a tiny kinetic mixing of a dark $U(1)_X^{}$ gauge field to the SM $U(1)_Y^{}$ gauge field. Specifically two vector-like fermions can carry the same $U(1)_Y^{}$ charge and the opposite $U(1)_X^{}$ charges of equal magnitude, as the $U(1)_X^{}$ and $U(1)_Y^{}$ gauge fields are assumed to respectively take an odd parity and an even parity under the $Z_2^{}$ symmetry. After the $Z_2^{}$ symmetry is spontaneously broken, the two vector-like fermions can obtain a small mass difference and then their contributions to the $U(1)_X^{} \times U(1)_Y^{}$ kinetic mixing can realize a large cancellation \cite{tg2026}. Moreover, these vector-like fermions are not allowed to leave an unacceptable relic density with electric charge. Instead, they can decay into the SM charged leptons with a TeV-scale complex dark scalar. The real and imaginary parts of this dark scalar can naturally obtain a mass split below the MeV scale after a dark Higgs scalar develops its vacuum expectation value (VEV) for generating a GeV-scale mass of the $U(1)_X^{}$ dark photon. The real and imaginary parts as two nearly degenerate dark matter particles thus can inelastically scatter off the nucleus in target through the dark photon mediation. Due to kinematic enhancement, the exothermic scattering rather than the endothermic scattering can explain the recently reported event in the LZ experiment since the two dark matter particles contribute the comparable relic densities in the present universe. Consequently the required interaction strength for interpreting the LZ signal can be lower enough to escape from the stringent limit from the solar capture.

\vspace*{2mm}
\section{Suppressed kinetic mixing}
\label{sec:kmixing}
\label{sec:2}
\vspace*{1mm}

As Abelian gauge fields, the dark $U(1)_{X}^{}$ gauge field $X_{\mu}^{}$ in principle can have a kinetic mixing with the SM $U(1)_Y^{}$ gauge field $B_\mu^{}$, i.e. 
\begin{eqnarray}
\label{kinetic1}
\mathcal{L}_K^{} &\supset &  -\frac{1}{4} X_{\mu\nu}^{} X^{\mu\nu}_{} -\frac{1}{4} B_{\mu\nu}^{} B^{\mu\nu}_{} -\frac{\epsilon}{2} X_{\mu\nu}^{} B^{\mu\nu}_{}\nonumber\\
[2mm]
&&\textrm{with}~~X_{\mu\nu}^{} = \partial_\mu^{} X_\nu^{} - \partial_\nu^{} X_\mu^{} ~~\textrm{and}~~ B_{\mu\nu}^{}= \partial_\mu^{} B_\nu^{} - \partial_\nu^{} B_\mu^{}\,.
\end{eqnarray}
However, it is allowed for these two Abelian gauge fields $X_{\mu}^{}$ and $B_{\mu}^{}$ to respectively carry an odd parity and an even parity under a $Z_2^{}$ discrete symmetry, i.e.
\begin{eqnarray}
\label{z21}
X_{\mu}^{} \stackrel{Z_{2}^{}}{\leftarrow\!\!\!-\!\!\!-\!\!\!-\!\!\!\rightarrow} -X_{\mu}^{}\,,~~B_{\mu}^{} \stackrel{Z_2^{}}{\leftarrow\!\!\!-\!\!\!-\!\!\!-\!\!\!\rightarrow} B_{\mu}^{}\,.
\end{eqnarray}
In consequence, although the first and second terms in the above Lagrangian remain invariant, the third term, i.e. the kinetic mixing term should be no longer available. Instead, we can introduce a real gauge-singlet Higgs scalar $\sigma$ with odd $Z_2^{}$-parity to construct a dimension-5 effective operator crossing the two gauge fields $X_{\mu}^{}$ and $B_{\mu}^{}$, i.e.
\begin{eqnarray}
\label{eff1}
\mathcal{L}_K^{} &\supset &  -\frac{\sigma}{2\Lambda} X_{\mu\nu}^{} B^{\mu\nu}_{} ~~\textrm{with}~~\sigma \stackrel{Z_{2}^{}}{\leftarrow\!\!\!-\!\!\!-\!\!\!-\!\!\!\rightarrow} -\sigma \,.\end{eqnarray}
After the real Higgs scalar $\sigma$ spontaneously develops its VEV, i.e.
\begin{eqnarray}
\label{breaking}
\sigma =v_\sigma^{} +  h_\sigma^{} 
\end{eqnarray}
the gauge fields $X_{\mu}^{}$ and $B_{\mu}^{}$ can obtain their kinetic mixing again. Here $v_\sigma^{}$ and $h_\sigma^{}$ are the VEV and the Higgs boson, respectively.

Note that the spontaneous breaking of the $Z_2^{}$ discrete symmetry should produce the so-called domain walls to dominate the energy density of the universe. In order to eliminate these domain walls promptly, we simply assume that this $Z_{2}^{}$ symmetry should be spontaneously broken before the inflationary epoch ends. For this purpose, we require the VEV $v_\sigma^{}$ larger than the reheating temperature $T_{R}^{}$, i.e. $v_\sigma^{} > T_{R}^{}$. For example, some inflation models allow $T_R^{} \lesssim 10^{6}\,\textrm{GeV}$ \cite{dkw2014,cdek2015}. Consequently, we can safely assume $v_\sigma^{} =\mathcal{O}\left(10^{7}_{}\,\textrm{GeV}\right)$.

Clearly the effective operator (\ref{eff1}) can be born in some renormalizable theories. As an example, we here seek help from the following two vector-like fermions as the twin partners under the present $Z_2^{}$ symmetry, i.e.
\begin{eqnarray}
\label{fermion1}
\psi_{1L,1R}^{}\left(+1,-1\right)   \stackrel{Z_{2}^{}}{\leftarrow\!\!\!-\!\!\!-\!\!\!-\!\!\!\rightarrow} \psi_{2L,2R}^{}\left(-1,-1\right)\,.
\end{eqnarray}
Here and hereafter the brackets following the fields describe the transformations under the $ U(1)_X^{} \times U(1)_Y^{}$ gauge groups. The relevant terms involving the vector-like fermions should include, 
\begin{eqnarray}
\label{kinetic1}
\mathcal{L}&\supset & i \bar{\psi}_{1L} \gamma^\mu_{} D_\mu^{} \psi_{1L}^{} + i \bar{\psi}_{1R} \gamma^\mu_{} D_\mu^{} \psi_{1R}^{}  + i \bar{\psi}_{2L} \gamma^\mu_{} D_\mu^{} \psi_{2L}^{} + i \bar{\psi}_{2R} \gamma^\mu_{} D_\mu^{} \psi_{2R}^{}  \nonumber\\
[2mm]
&&-y_\psi^{}\sigma \left[\left(\bar{\psi}_{1L}^{}\psi_{1R}^{} - \bar{\psi}_{2L}^{}\psi_{2R}^{}\right)+\textrm{H.c.} \right]- \bar{m}_{\psi}^{}\left[\left(\bar{\psi}_{1L}^{}\psi_{1R}^{} +\bar{\psi}_{2L}^{}\psi_{2R}^{} \right) +\textrm{H.c.}\right]\,,
\end{eqnarray}
where the covariant derivatives are given by 
\begin{eqnarray}
\label{covariant}
D_\mu^{} \psi_{1L,1R}^{}&=& \left(\partial_\mu^{} + i  g_X^{} X^{}_{\mu} - i    g' B^{}_{\mu} \right) \psi_{1L,1R}^{}\,,\nonumber\\
[2mm]
D_\mu^{} \psi_{2L,2R}^{}&=& \left(\partial_\mu^{} - i  g_X^{} X^{}_{\mu} - i    g' B^{}_{\mu} \right) \psi_{2L,2R}^{}\,,
\end{eqnarray}
with $g_{X}^{}$ and $g'$ being the gauge couplings. Because of their Yukawa couplings with the Higgs scalar $\sigma$ for the spontaneous $Z_2^{}$ symmetry breaking (\ref{breaking}), the vector-like fermions $\psi_{1}^{}= \psi_{1L}^{} + \psi_{1R}^{}$ and $\psi_{2}^{}= \psi_{2L}^{} + \psi_{2R}^{}$ can obtain a mass split, i.e. 
\begin{eqnarray}
\label{msplit}
m_{\psi_1}^{} = \bar{m}_{\psi}^{} +y_\psi^{} v_\sigma^{}~~\textrm{and} ~~ m_{\psi_2}^{} = \bar{m}_{\psi}^{}  - y_\psi^{} v_\sigma^{}~~\textrm{with}~~m_{\psi_1}^{} - m_{\psi_2}^{} =2 y_\psi^{} v_\sigma^{}\,.
\end{eqnarray}

 \begin{figure*}
\centering
\includegraphics[scale=0.75]{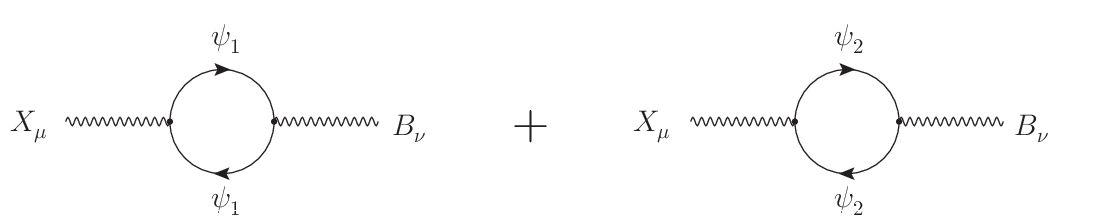} \caption{\label{1loop} The two vector-like fermions $\psi_{1}^{}$ and $\psi_2^{}$ can individually contribute to the $U(1)_X^{}\times U(1)_Y^{}$ kinetic mixing at one-loop level. Since these two fermions carry the same $U(1)_Y^{}$ charge and the equal but opposite $U(1)_X^{}$ charges, their one-loop contributions to this kinetic mixing should have a cancellation depending on their mass difference.}
\end{figure*}

As shown in Fig. \ref{1loop}, the two fermions $\psi_1^{}$ and $\psi_{2}^{}$ can individually mediate the $U(1)_X^{}\times U(1)_Y^{}$ kinetic mixing at one-loop level because they carry both $U(1)_X^{}$ and $U(1)_Y^{}$ charges. Remarkably, their $U(1)_Y^{}$ charges are exactly same while their $U(1)_X^{}$ charges are equal but opposite. Consequently their contributions to the $U(1)_X^{}\times U(1)_Y^{}$ kinetic mixing should be completely cancelled with each other as long as their masses are assumed exactly same. Actually this is just the reflection of the $Z_2^{}$ discrete symmetry. Now the two fermions $\psi_1^{}$ and $\psi_{2}^{}$ can have a mass split through the spontaneous breaking of this $Z_2^{}$ symmetry. There should be a nonzero difference between their contributions to the $U(1)_X^{}\times U(1)_Y^{}$ kinetic mixing. Specifically we can calculate 
\begin{eqnarray}
\label{c121}
\epsilon &=& c_1^{} + c_2^{} = \frac{g_X^{} g'}{12\pi^2_{}}   \ln \left( \frac{m_{\psi_1}^{2}}{m_{\psi_2}^{2}} \right)  \nonumber\\
[2mm]
&&\textrm{with} ~~ c_{1}^{}= -\frac{g_X^{} g'}{12\pi^2_{}} \ln\left(\frac{\mu^2_{}}{m_{\psi_1}^{2}} \right)\,, ~~ c_{2}^{}= +\frac{g_X^{} g'}{12\pi^2_{}} \ln\left(\frac{\mu^2_{}}{m_{\psi_2}^{2}}\right)\,.
 \end{eqnarray}
Here we have adopted the minimal subtraction scheme with $\mu$ being the renormalization scale. In the limiting case where the two vector-like fermions have a nearly degenerate mass spectrum, we can simplify the kinetic mixing (\ref{c121}) to be 
\begin{eqnarray}
\label{c122}
\epsilon \simeq  \frac{g_X^{} g'}{3\pi^2_{}}  \frac{y_\sigma^{} v_\sigma^{}}{\bar{m}_{\psi}^{}}~~\textrm{for}~~ m_{\psi_1}^{} - m_{\psi_2}^{} =2y_\psi^{} v_\sigma^{} \ll \bar{m}_{\psi}^{} \,. 
\end{eqnarray}

 \begin{figure*}
\centering
\includegraphics[scale=0.75]{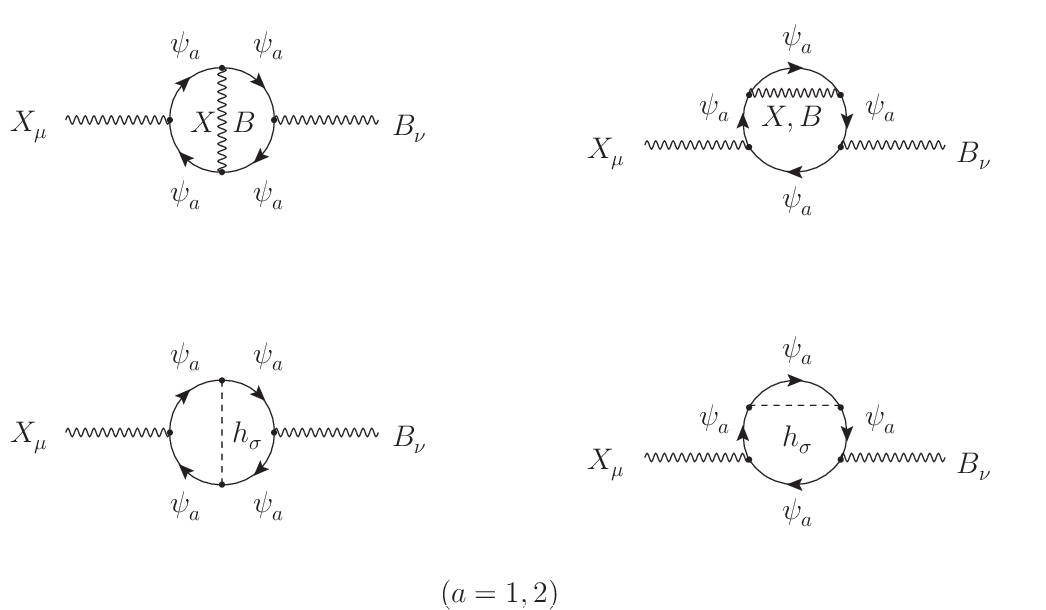} \caption{\label{2loop} The two vector-like fermions $\psi_{1}^{}$ and $\psi_2^{}$ can individually contribute to the $U(1)_X^{}\times U(1)_Y^{}$ kinetic mixing at two-loop level. Since these two fermions carry the same $U(1)_Y^{}$ charge and the equal but opposite $U(1)_X^{}$ charges, their two-loop contributions to this kinetic mixing should still have a cancellation depending on their mass difference.}
\end{figure*}

We also check the two-loop corrections. The relevant diagrams are shown in Fig. \ref{2loop}. We have 
\begin{eqnarray}
\label{c123}
\delta c_1^{} &=&-\frac{g_X^{}g' \left(g_X^2+ g'^2_{}\right)}{(16\pi^2_{})^2_{}}\left[\frac{55}{3}+ 4\ln\left(\frac{\mu^2_{}}{m_{\psi_1}^2}\right)\right] - \frac{g_X^{}g' y_\psi^2}{(16\pi^2_{})^2_{}}\left[\frac{11}{2}+ 2\ln\left(\frac{\mu^2_{}}{m_{\psi_1}^2}\right)\right] \,,\nonumber\\
[2mm]
\delta c_2^{} &=&+\frac{g_X^{}g' \left(g_X^2+ g'^2_{}\right)}{(16\pi^2_{})^2_{}}\left[\frac{55}{3}+ 4\ln\left(\frac{\mu^2_{}}{m_{\psi_2}^2}\right)\right] - \frac{g_X^{}g' y_\psi^2}{(16\pi^2_{})^2_{}}\left[\frac{11}{2}+ 2\ln\left(\frac{\mu^2_{}}{m_{\psi_2}^2}\right)\right]\,.
 \end{eqnarray}
The $U(1)_X^{}\times U(1)_Y^{}$ kinetic mixing then should be modified by 
\begin{eqnarray}
\epsilon &=& \frac{g_X^{} g'}{12\pi^2_{}}   \ln \left( \frac{m_{\psi_1}^{2}}{m_{\psi_2}^{2}} \right) +\delta c_1^{} + \delta c_2^{}=   \left\{\frac{g_X^{} g'}{12\pi^2_{}}  
+ \frac{2g_X^{}g' \left[2\left(g_X^2+ g'^2_{}\right)-y_\psi^2\right]}{(16\pi^2_{})^2_{}} \right\} \ln \left( \frac{m_{\psi_1}^{2}}{m_{\psi_2}^{2}} \right)  \nonumber\\
[2mm]
&\simeq& \left\{\frac{g_X^{} g'}{3\pi^2_{}}  
+ \frac{g_X^{}g' \left[2\left(g_X^2+ g'^2_{}\right)-y_\psi^2\right]}{32\pi^4_{}} \right\}  \frac{y_\psi^{} v_\sigma^{}}{\bar{m}_{\psi}^{}} \,. 
\end{eqnarray}
We hence can expect this kinetic mixing to arrive at an extremely small value. For example, we can obtain
\begin{eqnarray}
\epsilon = 4.2\times 10^{-9}_{}~~\textrm{for}~~g'=0.36\,,~g_X^{}=0.34\,,~y_\psi^{}=0.1\,,~v_\sigma^{}=10^{7}_{}\,\textrm{GeV}\,,~\bar{m}_{\psi}^{}=10^{12}_{}\,\textrm{GeV}\,.
\end{eqnarray}

\vspace*{3mm}
\section{Physical dark photon}
\label{sec:undp}
\label{sec:3}
\vspace*{1mm}

Due to the $U(1)_X^{}\times U(1)_Y^{}$ kinetic mixing, the $U(1)_X^{}$ gauge field $X_\mu^{}$ can couple to the SM fermions. To diagonalize the kinetic terms, we can consider the non-unitary transformation as follows, i.e.
\begin{eqnarray}
X_\mu^{} = \frac{1}{\sqrt{1-\epsilon^2}} A'^{}_\mu \,,~~B_\mu^{} = B'^{}_\mu -\frac{\epsilon}{\sqrt{1-\epsilon^2}} A'^{}_\mu \,.
\end{eqnarray}
In this new basis, we should rewrite the terms involving the neutral gauge bosons in the SM, i.e.
\begin{eqnarray}
\label{zxa}
\mathcal{L}&\supset& e J^{em}_{\mu}\left(A^\mu_{} - \bar{\epsilon } A'^\mu_{}\right)  +\frac{g}{\sin\theta_W^{}}J^0_\mu \left(Z^\mu_{} +\bar{\epsilon} \tan\theta_W^{} A'^\mu_{}\right)  \nonumber\\
[2mm]
&& +   \frac{1}{2} m_Z^2 \left(Z_\mu^{} + \bar{\epsilon} \tan\theta_W^{}  A'^{}_\mu\right) \left(Z^\mu_{} + \bar{\epsilon} \tan\theta_W^{}  A'^\mu_{}\right)~~\textrm{with} ~~\bar{\epsilon} =  \frac{\epsilon \cos\theta_W^{}}{\sqrt{1-\epsilon^2_{}}} \,.
\end{eqnarray}
Here $A_\mu^{}$ and $Z_\mu^{}$ are defined by 
\begin{eqnarray}
A_\mu^{}= W^{3}_\mu \sin\theta_W^{} + B'^{}_\mu \cos\theta_W^{}\,,~~   Z_\mu^{}= W^{3}_\mu \cos\theta_W^{} - B'^{}_\mu \sin\theta_W^{}\,,
 \end{eqnarray}
with $W^{3}_\mu$ being the third component of the $SU(2)_{L}^{}$ gauge fields and $\theta_W^{}$ being the Weinberg angle, i.e. $\sin^2_{}\theta_W^{}\simeq 0.231$. Moreover, $g$ is the $SU(2)_L^{}$ gauge coupling and $e$ is determined by $e= g \sin\theta_W^{}$. Finally, $J^{em}_\mu$ is the electromagnetic current, i.e.
\begin{eqnarray}
J^{em}_\mu &=& -\frac{1}{3}\left( \bar{d}\gamma_\mu^{} d + \bar{s}\gamma_\mu^{} s + \bar{b}\gamma_\mu^{} b\right) 
+ \frac{2}{3}\left(  \bar{u}\gamma_\mu^{}u + \bar{c}\gamma_\mu^{} c +\bar{t}\gamma_\mu^{} t \right)  
- \left(\bar{e}\gamma_\mu^{} e +\bar{\mu}\gamma_\mu^{} \mu +\bar{\tau}\gamma_\mu^{} \tau   \right)\,,~~
 \end{eqnarray}
while $J^{0}_\mu$ is the neutral current, i.e. 
\begin{eqnarray}
J^{0}_\mu &=& \frac{1}{2}\sum_{f}^{} \left[g_L^f  \bar{f}\gamma_\mu^{} \left(1-\gamma_5^{}\right) f  +g_R^f  \bar{f}\gamma_\mu^{} \left(1+\gamma_5^{}\right) f  \right] ~~\textrm{with}\nonumber\\
[2mm]
&&\begin{array}{lcl}
g_{L}^{\nu_e^{}} =g_{L}^{\nu_\mu^{}} =g_{L}^{\nu_\tau^{}} =\frac{1}{2}\,, &~~& g_{R}^{\nu_e^{}} =g_{R}^{\nu_\mu^{}} =g_{R}^{\nu_\tau^{}} =0\,;\\
[2mm]
g_L^{e}=g_L^{\mu}=g_L^{\tau}= -\frac{1}{2} +\sin^2_{}\theta_W^{}\,, &~~& g_R^{e}=g_R^{\mu}=g_R^{\tau} = \sin^2_{}\theta_W^{}\,;\\
[2mm]
g_L^{u} =g_L^{c} =g_L^{t} = \frac{1}{2}-\frac{2}{3}\sin^2_{}\theta_W^{}\,, &~~& g_R^{u} =g_R^{c} =g_R^{t} = -\frac{2}{3}\sin^2_{}\theta_W^{}\,;\\
[2mm]
g_L^{d}= g_L^{s}=  g_L^{b}=  -\frac{1}{2}+\frac{1}{3}\sin^2_{}\theta_W^{}\,, &~~& g_R^{d}= g_R^{s}=  g_R^{b} = \frac{1}{3}\sin^2_{}\theta_W^{}\,.
\end{array}
\end{eqnarray}

So far the dark photon $A'^{}_\mu$ remains massless. If the dark photon is expected to be massive, the corresponding dark gauge symmetry $U(1)_X^{}$ should be spontaneously broken. For this purpose, we can introduce a Higgs scalar as below,
\begin{eqnarray}
\label{darkhiggs}
\xi (+1,0)  \stackrel{Z_{2}^{}}{\leftarrow\!\!\!-\!\!\!-\!\!\!-\!\!\!\rightarrow}  \xi^\ast_{}(-1,0) ~~\textrm{with}~~  \xi=\frac{1}{\sqrt{2}} \left(v_\xi^{} + h_\xi^{}\right) \,.
\end{eqnarray}
The dark photon then can acquire its mass as below,
\begin{eqnarray}
\label{xx}
\mathcal{L}\supset \frac{1}{2} m_{A'}^2 A'^{}_\mu A'^{\mu}_{} ~~\textrm{with}~~m_{A'}^{2} &=& \frac{g_X^2 v_\xi^2}{1-\epsilon^2}
= \frac{4\pi \alpha_X^{} v_\xi^2}{1-\epsilon^2}\nonumber\\
[2mm]
& =&\left(1\,\textrm{GeV} \right)^2_{}\left(\frac{v_\xi^{}}{3\,\textrm{GeV}}\right)^2_{} \left(\frac{\alpha_X^{}}{0.0092}\right) \frac{1}{1-\epsilon^2}\,. 
\end{eqnarray}
Here and hereafter we conveniently define
\begin{eqnarray}
\alpha_X^{}=\frac{g_X^2}{4\pi} =0.0092 \left(\frac{g_X^{}}{0.34}\right)^2_{}~~\textrm{in~analogy with}~~\alpha=\frac{e^2_{}}{4\pi}\simeq 1/137\,.
\end{eqnarray}
Clearly the $A'^{}_\mu$ dark photon now has a mass mixing with the $Z_\mu^{}$ boson. The physical states can be given by 
\begin{eqnarray}
&&\hat{Z}_\mu^{}= Z_\mu^{}\cos\beta  + A'^{}_\mu \sin\beta \,,~~
\hat{A}'^{}_\mu=  A'_\mu\cos\beta- Z_\mu^{}\sin\beta\,,
\end{eqnarray}
with the mass eigenvalues,  
\begin{eqnarray}
m_{\hat{Z}}^{2} &=& \frac{1}{2}\left[m_{A'}^2 +(1+\bar{\epsilon}^2_{} \tan^2_{}\theta_W^{})m_Z^2 \right]  + \frac{1}{2} \sqrt{\left[ m_{A'}^2 +(1+\bar{\epsilon}^2_{} \tan^2_{}\theta_W^{})m_Z^2 \right]^2_{}- 4 m_{Z}^2 m_{A'}^2 }\,,\nonumber\\
[2mm]
m_{\hat{A}'}^{2} &=&  \frac{1}{2}\left[m_{A'}^2 +(1+\bar{\epsilon}^2_{} \tan^2_{}\theta_W^{})m_Z^2 \right] - \frac{1}{2} \sqrt{\left[ m_{A'}^2 +(1+\bar{\epsilon}^2_{} \tan^2_{}\theta_W^{})m_Z^2 \right]^2_{}- 4 m_{Z}^2 m_{A'}^2 }\,,\nonumber\\
[2mm]
&&
\end{eqnarray}
and the rotation angle, 
\begin{eqnarray}
\tan 2\beta = \frac{2 \bar{\epsilon} \tan \theta_W^{}m_{Z}^2 }{ \left(1-\bar{\epsilon}^2_{}\tan^2_{}\theta_W^{}\right)m_Z^2-m_{A'}^2}\,.
\end{eqnarray}

In some limiting cases, we can consider the simplifications as below,
\begin{eqnarray}
Z_\mu^{} &\simeq & \hat{Z}_\mu^{} - \bar{\epsilon}\tan\theta_W^{} \left(1+\frac{m_{{A}'}^{2}} {m_{Z}^2} \right)\hat{A}'^{}_\mu~~\textrm{with}~~m_{\hat{Z}}^{2} \simeq m_Z^2 \simeq \left(91.2\,\textrm{GeV}\right)^2_{}\,, \nonumber\\
[2mm]
A'^{}_\mu &\simeq& \hat{A}'^{}_\mu + \bar{\epsilon}\tan\theta_W^{} \left(1+\frac{m_{{A}'}^{2}} {m_{Z}^2} \right)  \hat{Z}_\mu~~\textrm{with}~~m_{\hat{A}'}^{2} \simeq m_{A'}^2~~\textrm{for}~~\bar{\epsilon} \ll 1\,, ~~m_{A'}^2\ll m_{Z}^2\,.
\end{eqnarray}
Accordingly, the current interactions in Eq. (\ref{zxa}) can be well approximated to
\begin{eqnarray}
\label{zxa4}
\mathcal{L}&\supset& e J^{em}_{\mu}\left(A^\mu_{}    - \bar{\epsilon }   \hat{A}'^\mu_{} -  \bar{\epsilon}^2_{} \tan\theta_W^{}  \hat{Z}^\mu_{}   \right) +\frac{g}{\sin\theta_W^{}}J^0_\mu \left( \hat{Z}^\mu_{}  - \frac{\bar{\epsilon} \tan\theta_W^{}m_{A'}^2}{m_{Z}^2}  \hat{A}'^\mu_{}    \right) \,.
\end{eqnarray}

\vspace*{3mm}
\section{Inelastic dark matter}
\label{sec:idm}
\label{sec:4}
\vspace*{1mm}

In the previous demonstration in Sec. \ref{sec:kmixing}, the two vector-like fermions $\psi_{1,2}^{}$ carry an electric charge $-1$. Therefore, their relic density can not be accepted at all. This means they should have proper decay channels. For this purpose, we introduce a complex dark scalar with $U(1)_X^{}$ charge, i.e.
\begin{eqnarray}
\chi (+1,0) =\frac{1}{\sqrt{2}}\left(\chi_{\textrm{R}}^{} +i \chi_{\textrm{I}}^{}\right) \stackrel{Z_{2}^{}}{\leftarrow\!\!\!-\!\!\!-\!\!\!-\!\!\!\rightarrow} \chi^{\ast}_{}(-1,0) = \frac{1}{\sqrt{2}}\left(\chi_{\textrm{R}}^{} - i \chi_{\textrm{I}}^{}\right) \,,
\end{eqnarray}
which should have the following kinetic term,
\begin{eqnarray}
\label{gaugechi}
\mathcal{L}&\supset& \left(D_\mu^{}\chi \right)^\dagger_{} \left(D^\mu_{}\chi\right)= \left(\partial_\mu^{} \chi  + i g_X^{} X_\mu^{} \chi \right)^\ast_{} \left(\partial^\mu_{} \chi  + i g_X^{} X^\mu_{} \chi \right) \nonumber\\
[2mm]
&=& \frac{1}{2}\left(\partial_\mu^{}\chi_{\textrm{R}}^{}\right) \left(\partial^\mu_{}\chi_{\textrm{R}}^{}\right) +  \frac{1}{2}\left(\partial_\mu^{}\chi_{\textrm{I}}^{}\right) \left(\partial^\mu_{}\chi_{\textrm{I}}^{}\right) +\frac{1}{2} g_X^2 \chi_{\textrm{R}}^{2} X_\mu^{} X^\mu_{} +\frac{1}{2} g_X^2 \chi_{\textrm{I}}^{2} X_\mu^{} X^\mu_{}\nonumber\\
[2mm]
&&+\frac{i}{2} g_X^{} \chi_\textrm{I}^{} \left(\partial_\mu^{}\chi_{\textrm{R}}^{}\right) X^\mu_{}   - \frac{i}{2} g_X^{} \chi_\textrm{R}^{} \left(\partial_\mu^{}\chi_{\textrm{I}}^{}\right) X^\mu_{}  \,.
\end{eqnarray}
We hence can have the Yukawa couplings of the vector-like fermions $\psi_{1,2}^{}$ and the dark scalar $\chi$ to the SM right-handed charged leptons, i.e.
\begin{eqnarray}
\mathcal{L}&\supset& - \sum_{l=e,\mu,\tau}^{} \left[f_{l}^{} \left(\chi \bar{\psi}_{1L} l_R^{} + \chi^\ast_{} \bar{\psi}_{2L}^{} l_R^{} \right)+\textrm{H.c.}   \right] \,.
\end{eqnarray}

The dark scalar $\chi$ in principle can spontaneously develop a VEV, like the dark Higgs scalar $\xi$ introduced in Eq. (\ref{darkhiggs}). In this case, the vector-like fermions $\psi_{1,2}^{}$ can mix with the SM charged leptons. Depending on the scale of their own masses and the structure of their Yukawa couplings with the SM charged leptons, the vector-like fermions $\psi_{1,2}^{}$ can safely escape from the existing experimental constraints. Alternatively, we can prevent the dark scalar $\chi$ from any nonzero VEVs by additionally imposing a global or discrete symmetry unbroken at any scales. For example, we consider the $Z_2^{}$ discrete symmetry as below, 
\begin{eqnarray}
Z_2^{\textrm{D}}: ~~ \psi_{1,2}^{}(-)\,,~~\chi (-) \,,~~\xi(+)\,.
\end{eqnarray}
Under this $Z_2^{\textrm{D}}$ symmetry, the renormalizable portal between the two dark scalars $\chi$ and $\xi$ should only contain the following two terms, i.e 
\begin{eqnarray}
\label{portchi}
V&\supset& \lambda \xi^\ast_{}\xi \chi^\ast_{}\chi + \frac{1}{4}\kappa \left[\left(\xi^\ast_{}\chi\right)^2_{}+\textrm{H.c.}\right]\,.
\end{eqnarray}
Clearly, the real and imaginary parts of the complex dark scalar $\chi$ can acquire a mass split after the dark Higgs scalar $\xi$ develops its VEV as shown in Eq. (\ref{darkhiggs}), i.e. 
\begin{eqnarray}
&& m_{\chi_{\textrm{R}}^{}}^2- m_{\chi_{\textrm{I}}^{}}^2 =  \frac{1}{4} \kappa v_\xi^2\,,  ~~ m_\chi^{} = \frac{1}{2}\left(m_{\chi_{\textrm{R}}^{}}^{}+m_{\chi_{\textrm{I}}^{}}^{}\right)\,, \nonumber\\
[2mm]
&& \delta = \left|m_{\chi_{\textrm{R}}^{}}^{} - m_{\chi_{\textrm{I}}^{}}^{} \right|  = \frac{\left|\kappa v_\xi^2\right|}{8 m_\chi^{}}=600\,\textrm{keV} \left(\frac{|\kappa|}{0.42}\right)\left(\frac{v_\xi^{}}{3\,\textrm{GeV}}\right)^2_{}\left(\frac{800\,\textrm{GeV}}{m_\chi^{}}\right)\,.\end{eqnarray}
Without loss of generality and for convenience, we denote the heavier and lighter states by 
\begin{eqnarray}
\chi_1^{}~~\textrm{with}~~m_{\chi_1^{}}^{}= \min\left \{m_{\chi_{\textrm{R}}^{}}^{}, m_{\chi_{\textrm{I}}^{}}^{}\right\}\,,~~\chi_2^{}~~\textrm{with}~~m_{\chi_2^{}}^{}= \max\left \{m_{\chi_{\textrm{R}}^{}}^{}, m_{\chi_{\textrm{I}}^{}}^{}\right\}\,.\end{eqnarray}
As long as the kinematics is allowed, one heavier state $\chi_2^{}$ can decay into one lighter state $\chi_1^{}$ with one dark photon or with two SM fermions before the BBN. In this case, the lighter and hence stable state $\chi_1^{}$ can be the unique dark matter particle. Otherwise, we should have two dark matter particles: a long-lived dark matter particle, i.e. the heavier state $\chi_2^{}$, and a stable dark matter particle, i.e. the lighter state $\chi_1^{}$. Actually we find 
\begin{eqnarray} 
\Gamma_{\chi_2^{}}^{} &=& \sum_{\alpha=e,\mu,\tau}^{}\Gamma\left(\chi_2^{}\longrightarrow \chi_1^{} + \nu_{L\alpha}^{}+\bar{\nu}_{L\alpha}^{} \right) \simeq  \frac{ 2\, \alpha_X^{}\alpha }{5\pi} \left(\frac{\bar{\epsilon}\, m_{A'}^2 }{m_{Z}^2 \sin2 \theta_W^{}}\right)^2_{} \frac{\delta^5_{}}{m_{\chi}^4}    \nonumber\\
[2mm]
&=& 5.8\times 10^{-49}_{}\,\textrm{eV}\left(\frac{\bar{\epsilon}}{4.2\times 10^{-9}_{}}\right)^2_{}  \left(\frac{\alpha_X^{}}{0.0092}\right) \left(\frac{800\,\textrm{GeV}}{m_\chi^{}}\right)^4_{}\left(\frac{m_{A'}^{}}{1\,\textrm{GeV}}\right)^4_{} \left(\frac{\delta}{600\,\textrm{keV}}\right)^{5}_{} \nonumber\\
& \ll& 10^{-33}_{}\,\textrm{eV}\,.
\end{eqnarray}
Therefore, the heavier and lighter dark matter should coexist with comparable relic density in the present universe.

The two dark scalars $\chi_{1,2}^{}$ with a small mass split can annihilate and co-annihilate into the light species through the gauge interactions (\ref{gaugechi}) and the Higgs portal interactions (\ref{portchi}), i.e.
\begin{eqnarray}
\chi+\chi^\ast_{}\rightarrow \hat{A}'+\hat{A}'\,,~~\chi+\chi^\ast_{}\rightarrow \xi+\xi^\ast_{}\,,~~\chi+\chi \rightarrow \xi+\xi\,,~~\chi^\ast_{}+\chi^\ast_{} \rightarrow \xi^\ast_{}+\xi^\ast_{}\,.
\end{eqnarray}
The thermally-averaged cross section can be computed by 
\begin{eqnarray}
\label{ann}
\langle\sigma_{\textrm{A}}^{} v_{\textrm{rel}}^{} \rangle&\simeq &\left( \pi \alpha_X^2+\frac{2 \lambda^2_{}+\kappa^2_{}}{128\pi}\right)\frac{1}{ m_\chi^2} \,.
\end{eqnarray}
Thus the dark matter relic density can be described by  
\begin{eqnarray}
\label{relic}
\Omega_{\chi}^{} h^2_{} \simeq \frac{0.1\,\textrm{pb}}{\langle\sigma_{\textrm{A}}^{} v_{\textrm{rel}}^{} \rangle}\,,
\end{eqnarray}
when the dark matter mass is roughly in the range from a few GeV to a few TeV. In order to obtain the measured value of the dark matter relic density, we can take a proper parameter choice such as the following example, 
\begin{eqnarray}
\label{ann3}
\langle\sigma_{\textrm{A}}^{} v_{\textrm{rel}}^{} \rangle&\simeq & 1\,\textrm{pb} \left(\frac{800\,\textrm{GeV}}{m_\chi^{}}\right)^2_{}~~\textrm{for}~~\alpha_X^{}\simeq 0.0092\,,~~\lambda\simeq 0.44\,,~~\textrm{and}~~\kappa\simeq 0.42\,.
\end{eqnarray}

On the other hand, the LZ event is expected to come from the inelastic scattering of dark matter off nucleus. In the presence of an isospin symmetry between proton and neutron, we can fit the LZ  signal by the three parameters as follows \cite{fhwz2026},
\begin{eqnarray}
\label{dmnuleus}
m_\chi^{}\,,~~\delta\,,~~\sigma_{N}^0=A^2_{} \frac{\mu_N^2}{\mu_p^2}\sigma_p^{}\,.
\end{eqnarray}
Here $A$ represents the total number of nucleons in the Xenon nucleus, $\sigma_p^{}$ is the spin-independent dark-matter-proton scattering cross section, $m_\chi^{}$ and $\delta$ denote the dark matter mass and the mass split, while $\mu_p^{}$ and $\mu_N^{}$ are the reduced masses, i.e. 
\begin{eqnarray}
\mu_p^{}=\frac{m_p^{} m_\chi^{} }{m_p^{}+m_\chi^{}}\,,~~\mu_N^{}=\frac{m_N^{} m_\chi^{} }{m_N^{}+m_\chi^{}}\,.
\end{eqnarray}
In the present model, the neutron should be almost absent from the inelastic dark matter scattering because it has no significant vector-current interaction with the dark photon. We hence should modify the dark-matter-nuclei cross section in Eq. (\ref{dmnuleus})  to be 
\begin{eqnarray}
\sigma_{N}^0 &=&Z^2_{} \frac{\mu_N^2}{\mu_p^2}\sigma_p^{} \,,
\end{eqnarray}
with $Z$ being the nuclear charge number. Here the dark-matter-proton scattering cross section is determined by 
\begin{eqnarray}
\sigma_p^{} &=& \frac{16\pi\,\bar{\epsilon}^2_{} \alpha_X^{} \alpha \mu_p^2}{m_{\hat{A}'}^4} = 2\times  10^{-44}_{}\,\textrm{cm}^2_{} \left(\frac{\bar{\epsilon}}{4.2\times 10^{-9}_{}}\right)^2_{}\left(\frac{\alpha_X^{}}{0.0092}\right)\left(\frac{\mu_p^{}}{m_p^{}}\right)^{2} \left(\frac{1\,\textrm{GeV}}{m_{\hat{A}'}^{}}\right)^4_{}\,.\nonumber\\
[2mm]
&&
\end{eqnarray}
Now the heavier and lighter dark matter particles contribute the comparable relic densities in the present universe, so that their exothermic inelastic scattering should always induce the stronger signal than their endothermic inelastic scattering due to kinematic enhancement. This means the interaction strength for interpreting the LZ event can be lower enough to evade the IceCube constraint. For this purpose, we can take  the following inputs,  
 \begin{eqnarray}
\sigma_p^{} = 2\times  10^{-44}_{}\,\textrm{cm}^2_{}\,,~~\delta= 600\,\textrm{keV} \,,~~m_\chi^{}=800\,\textrm{GeV}\,,
\end{eqnarray}
as an example suggested in \cite{fhwz2026}. 
 
\vspace*{3mm}
\section{Conclusion}
\label{sec:con}
\label{sec:4}
\vspace*{1mm}

In this paper, we have shown that the dark $U(1)_X^{}$ gauge field can acquire a tiny kinetic mixing with the SM $U(1)_Y^{}$ gauge field through the spontaneous breaking of a $Z_2^{}$ discrete symmetry, under which the $U(1)_X^{}$ and $U(1)_Y^{}$ gauge fields respectively take an odd parity and an even parity. Specifically, we introduce two vector-like fermions, which carry the same $U(1)_Y^{}$ charge and the opposite $U(1)_X^{}$ charges of equal magnitude. After the $Z_2^{}$ symmetry is spontaneously broken, the two vector-like fermions can obtain a small mass difference to largely cancel their contributions to the $U(1)_X^{} \times U(1)_Y^{}$ kinetic mixing. Because these vector-like fermions are forbidden from leaving a relic density with electric charge, they can be expected to decay into the SM charged leptons with a complex dark scalar. The real and imaginary parts of this dark scalar can obtain a mass split below the MeV scale after a dark Higgs scalar develops its VEV for generating a GeV-scale mass of the $U(1)_X^{}$ dark photon. The real and imaginary parts thus can have an inelastic scattering off nuclei through the dark photon mediation. Remarkably, the exothermic dark matter scattering rather than the endothermic dark matter scattering can explain the recently reported event in the LZ experiment due to kinematic enhancement. The required interaction strength for interpreting the LZ event thus can be lower enough to satisfy the stringent constraints from the IceCube neutrino searches.

\vspace*{5mm}
\noindent
{\bf\large Acknowledgements}
\\[1.5mm]

This work was supported in part by the National Natural Science Foundation of China under Grant No. 12175038. 


\baselineskip 17pt

\vspace{5mm}
%

\end{document}